\documentclass[12pt,preprint]{aastex}

\newcommand{\wga}{1WGA~J0447.9-0322~}

\begin{document}

\title{ Optical variability of the strong-lined and X-ray bright source\\
1WGA J0447.9-0322}

\author{R. Nesci, M. Mandalari and S. Gaudenzi}
\affil{ Department of Physics, University of Roma La Sapienza, \\
Piazzale A. Moro 2, I-00185 Roma, Italy 
}
\email{roberto.nesci@uniroma1.it}

\begin{abstract}

We present the historic light curve of 1WGA J0447.9-0322, 
spanning the time interval from 1962 to 1991, built using the Asiago 
archive plates. The source shows small fluctuations of about 0.3 mag around
B=16 until 1986 and a fast dimming of its average level by about 0.5 mag after that date, again with small short term variations.
The variability pattern is within the values shown by other QSOs with long
term monitoring, notwithstanding its high X-ray/optical ratio.
We present also its overall SED using literature data and recent UV-optical SWIFT observations.
\end{abstract}



\section{Introduction}

The DXRBS and RGB samples of blazars (Perlman et al. 1998;
Laurent-Muehleisen et al.
1999) discovered a number of Flat
Spectrum Radio Quasars (FSRQ) with strong X-ray fluxes (Padovani et al.
2003). This
finding was in contrast with the previous expectation that strong-lined
Quasars
should not be bright X-ray emitters.

The so-called "blazar sequence", proposed by Fossati et al. (1998), indeed 
showed a marked
correlation between the peak
frequency of the spectral energy distribution, in the radio to soft X-ray
range, 
and the X-ray/optical flux ratio.
According to that sequence, the peak frequency of the low-energy branch of
the
spectral energy distribution is inversely correlated to the bolometric
power of the
source, so that only intrinsically faint sources should be capable of
accelerating
electrons to the high energies required to push the syncrotron
emission up to the X-ray band: these sources are the so called HBL objects.

Very little information is
presentely available on these X-ray strong FSRQ, as they have been
discovered
relatively recently. One of these is the radio source PMN
J0447-0322 (Griffith et al. 1995), identified as an X-ray source in the
WGA catalogue (White et al. 1995, 2000). It is present as a bright source
in the Rosat All
Sky Survey (Voges et al. 1999) and as such it was included in ROSAT-based
catalogues of AGNs like the REX survey (Caccianiga et al. 2000). On the
basis of its optical and radio spectrum it was classified as a bright Flat
Spectrum
Radio Quasar (FSRQ) by Caccianiga et al. (2000) with a redshift z=0.773. It
is also
present in the 2MASS infrared survey (Barkhouse \& Hall 2001).

Despite its relative brightness (B$\sim$ 16) the optical variability of
\wga was not
studied in any detail: we made therefore a search in the Asiago plate
archive for
historical images and found a large number of useful Schimdt plates,
covering the
period from 1961 to 1985. 

In this paper we present the first optical historic light curve of \wga
and compare it with long term light curves of QSOs
taken from the literature to see if it is statistically different from that
of X-ray
faint sources.
We report also the results of a recent SWIFT observation of this source in
the
optical and X-ray bands and report an updated Spectral Energy Distribution.

\section{The 67/92 cm Photographic Material and Data Analysis}

We found 266 useful plates in the Asiago Observatory archive containing our
source: 
205 of them were obtained with the 50/40/120 cm Schmidt telescope (S40) and
61 with
the 90/67/245 cm one (S67). Several filter/emulsion combinations were used
over the
years, characterized by different effective wavelengths: 103aO+GG13
(closely matching
the Johnson's B filter), 103aO (3600-5000 \AA), TriX (3600-6700 \AA) and
Panchro
Royal (3800-6400 \AA). All these plates were taken as part of the Supernova
sky
patrol of the Asiago Observatory ($\mu ~Eri$ field) and were never used
before for the
study of \wga.
The covered time interval is from September 1962 to January 1991, but with
a highly
uneven sampling. Indeed the source can be properly observed from Asiago
only during the winter
because of its low declination. 

The plates were digitized at the Asiago Observatory with an EPSON 1680 Plus
scanner as part of a national program of digitization of astronomical plate
archives
(Barbieri et al. 2003). 
A sampling step of 16 micron (1600 dpi) was used, in grayscale/transparency
mode and
16 bit resolution. Plate scanning included also the unexposed borders to
measure the
plate fog level ($F$). Due to the presence of a residual scattered light in
the
scanner, we evaluated the instrumental zero value for each plate ($Z$)
using the
central pixels of the most overexposed
stars. The transformation of the recorded plate transparency $T$ of each
pixel into a
relative intensity $I$ was obtained applying the simple relation
$I=(F-Z)/(T-Z)$.
As there is no photometric sequence in the sky area near \wga we used the 
$B$
magnitudes of the GSC2 catalogue for our measures of \wga. To this purpose
we
selected
18 stars around the source covering the range $14.3 \le B \le 16.8$: the
faint
extreme is nearly the limit of detection for the small Schmidt telescope.

We checked that the selected stars were not variable and that their $B-R$
color
indices were similar to minimize any possible color effect. In this respect
\wga is
remakably bluer (B-R$\sim$0.5) than all our reference stars (B-R$\sim$1.0).
A finding chart of the reference stars is given in Fig.1, where  \wga
is marked with W. 
Table 1 gives for each star RA and DEC (J2000) taken from the GSC2
(column 1 and 2), the GSC2 identification (column 3), the R and B
magnitudes (columns
4 and 5), our adopted $B$ magnitude and internal error (column 6, discussed
below)
and its flag on the finding chart (column 7).

We started our search with the plates of the 67 cm Schmidt, which allows a
higher
photometric accuracy: nearly all the plates were taken with the 103aO
emulsion
without filter and have a limiting magnitude $B\sim$18.  Instrumental
magnitudes were
obtained with IRAF/apphot using a photometric aperture of 3 pixels (4.5
arcsec). The
scatter plot between these magnitudes and those of the GSC2 for each plate
showed
that a linear fit was quite satisfactory.
 The slopes of the fitting lines were always
between 0.7 and 1.4, as expected for the linearized response of a
photographic plate.
 The formal error of the $B$ magnitudes in the GSC2 catalogue is 0.4
mag, mainly due to the uncertainty of the zero
point calibration. Having 67 plates of the field we performed an
intercalibration of
the magnitudes of our reference stars and obtained the data listed in
column 6 of
Table 1. The differences with respect with the GSC2 $B$ values are in all
cases smaller than 0.1 mag, indicating that the GSC2 catalogue can be very
useful for
the construction of historical photographic light curves.

The magnitudes of column 6 (which we will call B/Asiago) were then used to
rebuild the calibration curve for each plate, which were again well fitted by a
linear relation with an appreciably smaller scatter. 
The $B$ magnitude of \wga in each plate was then derived from the
instrumental one using the relevant calibration curve. These magnitudes and 
the rms deviation of the fit are reported in Table 2 (electronic edition only), column 3 and 4. In several
cases, due to statistical fluctuations and to different plate quality, the rms
deviation is smaller than the errors of the individual reference stars given in
Table 1, which are derived from the whole plate set. The typical value is 0.1 mag
and is representative, in our opinion, of the actual photometric accuracy even
for good plates with smaller formal error.
The plate number and JD-2,400,000 are reported in column 1 and 2 of Table 2.

We checked if the color difference between \wga and the reference stars may
produce systematic effects in our photometry. We made a numerical simulation,
assuming a spectral shape
$F_{\nu}=A \times \nu^{-\alpha}$ and integrating the flux with the B filter
passband, normalized at the central wavelength. The
$\alpha$ values corresponding to the color index of our source (B-R=0.5) and of
our average reference stars (B-R=1.0) are 0.4 and 1.5 respectively.
Changing the spectral slope $\alpha$ by this amount gives a magnitude
variation of just 0.03 mag, which is below our typical photometric uncertainty.

Actually many BL Lac objects are "bluer when brighter", with a spectral slope
variation range of $\sim$0.5 (see e.g. Vagnetti et al. 2003; Fiorucci et al.
2004). Also if we consider \wga as a QSO rather than a BL Lac, the spectral index
variability has a limited range ($\sim$0.5, Trevese \& Vagnetti 2002).
From our simulation such a variation has a negligible effect (0.01 mag) on the
observed magnitude of our source, so that we can ignore the effect of a possible
color change with luminosity of \wga.

In no case was \wga fainter than our faintest reference stars, so
that our magnitudes are always derived by interpolation in the calibration
curve. The
resulting
light curve is plotted in Fig.2, together with the light curve of the
reference star r
(B=16.0) shifted 0.5 mag upwards for clarity.

\section{Digital Sky Survey data}
Two additional historical points were obtained from the digitized POSS-I
(date 1955-12-14, JD
2435455) and UKST DSS-II (date 1982-12-12, JD 2445315) blue plates
retrieved from the
Space Telescope Science Institute archive
(http://archive.stsci.edu/cgi-bin/dss\_plate\_finder),
using the same reference stars and data reduction procedure used for the
67cm Schmidt
plates.

The POSS-I plate gave a good linear fit for the instrumental magnitudes of
the
reference stars (rms=0.09) with $B$=15.85 for \wga (circle at JD 35455 in
Fig.2).
The UKST plate is that used for the construction of the GSC2 catalogue: in
our B/Asiago
magnitude scale we measured \wga at $B$=16.03 (circle at JD 45315 in
Fig.2): we have a
nearly
simultaneus observation in our database (plate S67-11792) at $B$=15.86, in
reasonable
agreement.

We found also two red plates in the STScI archive, a POSS-I plate (103aE
emulsion) made the
same
day as the blue one, and a UK Schmidt (IIIaF emulsion)
plate taken on 1995-11-14. We measured instrumental magnitudes of our
reference stars for
these
plates too (with IRAF/apphot) and made calibration curves
using the GSC2 red magnitudes. A linear fit in the magnitude range
15.4-16.0 was quite
satisfactory
with rms scatter $\sim$0.1 .
For both plates we get $R$=15.60, very near to the value $R$=15.63 given by
the GSC2
catalogue.
From the simultaneous 1955 B and R plates, the $B-R$ color index of \wga
was 0.27: assuming
that
this value is valid also for the 1995 we get that
the source in 1995 was as bright as in 1955 ($B$=15.85, star at JD 50036 in
Fig.2),
appreciably
brighter than the last S67 Asiago point in 1991 at $B$=16.3.

\section{SWIFT-UVOT observation of \wga}

\wga was observed several times by SWIFT (Gehrels et al. 2004) in the
framework of a program
of Blazar
monitoring, but only in one pointing
(2005-04-15) UVOT data were obtained. The observing strategy of the
satellite was to
get several short exposures in each of the 6 filters (V, B, U, UW1, UM2,
UW2):
to improve the S/N ratio we summed all the frames relative to each filter
and made
the
aperture photometry of our reference stars (and \wga) with the task
UVOTSOURCE in the
UVOT data
analysis package using a 6 arcsec radius for the V,B,U filters and 12
arcsec for the
remaining ones, as reccomended in the most recent cookbook (Breeveld
et al. 2005). The resulting magnitudes for each filter, not corrected for
the
galactic reddening, are collected
in column 2 of Table 3; in the same Table we show also the $\nu F_{\nu}$
values (in erg s$^{-1}$ cm$^2$), corrected for galactic extinction
assuming E(B-V)=0.05, which we will use in Section 6 to build the Spectral
Energy Distribution
(SED) of the source.

We made a consistency check for the $B$ filter between the UVOT and our
B/Asiago
magnitudes. To this purpose we made aperture photometry with UVOTSOURCE of
13
reference stars which fall within the UVOT field of view (star s was
excluded because it was
very near to the saturation limit of UVOT).
The comparison of the UVOT and B/Asiago magnitudes gave a best fit slope
1.10
with rms 0.04. 
The magnitude differences between the reference stars, due to the different
zero points
and slope, are not large, being 0.1 at $B$=14.7 and 0.3 at $B$=16.7, in any
case
within the formal error of the GSC2 absolute calibration.

We tried to further check this point observing the field of \wga at the 182
cm telescope of
Cima Ekar (Asiago) on 2006-01-04 with the CCD camera of the AFOSC
instrument, but the night
was not
photometric, so we could not check the zero point of the GSC2 magnitude
scale. Due to the
limited
field of view (8.8') of AFOSC we were able to include only 5 reference
stars (excluding star s
as
above), covering however the magnitude range of the \wga light curve. A
comparison of the AFOSC instrumental magnitudes with those of UVOT showed
that the linearity of the UVOT magnitude scale is rather good (slope=1.034,
rms 0.018 );
a small non-linearity of the UVOT instrument at the 5\% level was also
found by Li et
al. (2006) in the analysis of the stars in two supernova fields and
by some of us in the analysis of the reference stars in the fields of BL
Lacertae
(Tosti et al., 2006). 
We conclude therefore that the GSC2 scale is somewhat compressed.
This small compression has anyway little effect on the light curve
of \wga which has a maximum amplitude of only 1 mag: the peak to peak
systematic difference is
0.1 mag, comparable with the statistical error of our photographic
magnitudes. Therefore we
did not correct our light curve for this effect.

The magnitude of \wga observed by UVOT was $B$=16.60 on the
B/Asiago scale (open square in Fig.2): the same value was given by the
AFOSC observation 8
months later(filled square in Fig.2).
The source was therefore in the last years at a flux level similar to the
low state of the
years
1987-89 of the Asiago historic light curve.

\section{The Small Schmidt Plates}

Finally we made the photometry of our source also on the S40 plates, using
our
 B/Asiago values for the reference stars: due to the smaller plate scale
(206
arcse/mm) we adopted a photometric aperture radius of 2 pixels (6.6
arcsec).
The rms deviation of the linear fit between
nominal and instrumental magnitudes of the reference stars was higher than
for the 67
cm Schmidt, as expected for a smaller instrument. The scatter of the
reference
stars magnitudes around their nominal value is however well-behaved, i.e.
increases
monotonically from 0.08 mag at $B \sim$14.5 to 0.26 mag at $B \sim$16.5. 
In these plates the fainter stars of the sequence were rather near to the
plate
limit.
We decided
to use these lower quality plates just to look for possible large amplitude
flares
of \wga. The S40 plates
are indeed much more numerous that the S67 ones and allow therefore a
better
time coverage of the light curve, though with a worse photometric accuracy.

In a large number of nights, two consecutive plates were taken at the S40
with different emulsions: no systematic differences in the magnitudes of
\wga were
found between Panchro Royal
and 103aO, while a systematic difference of 0.23 mag was found between
Tri-X and
103aO, the Tri-X magnitudes being fainter. This is probably due to the
better violet
sensitivity of the 103aO emulsion, which is relevant for an object
definitely bluer
than the reference stars. We reported all the Tri-X magnitudes to the zero
point of
the 103aO emulsion, to be consistent with the light curve obtained with the
S67 and then we
averaged the data referring to the same night. The resulting
the light curve is shown in Fig.3. Large circles show the S67 points, while
the small
crosses show the values for the S40: error
bars are omitted for clarity.
The overall behaviour of the light curve is similar to that of the S67, but
with a substantial
scatter. The remarkable point is that
no strong flares were detected. The source was never brighter than
$B$=15.5.
The list of all the $B$ magnitudes and relative
uncertainties of \wga is given in Table 2 (electronic edition only): only one plate number is listed
when two plates were averaged.

\section{The Spectral Energy Distribution}

A spectral energy distribution of \wga was published by Padovani et al
(2003) using
literature data. An enhanced version, including JHK magnitudes taken from
the 2MASS
(Cutri et al. 2003) and preliminary X-ray and optical data from XMM-Newton
was
published by Landt et al. (2005). The source was not detected by EUVE.

The SED of the source is shown in Fig.\ref{sed}, including literature
radio, JHK, 
GSC2 data and
our optical and UV data from SWIFT. The radio point 
at 4.8 GHz is from the PMN survey on November 1990 (Griffith et al . 1995)
and is nearly
simultaneously to the ROSAT observation used in
the WGA catalogue. The X-ray flux is taken from the NED, and we checked the
result converting the WGA count-rate to flux using
the PIMMS tool at HEASARC. The observation at 1.4 GHz is given by the NVSS
(Condon
et al. 1998; 87 $\pm$ 3 mJy) and shows no appreciable polarization.
From our light curve the B magnitude at the epoch of the ROSAT and PMN
observations was about
16.3, only 0.2 mag fainter than the GSC2 point. The high X-ray/optical flux
ratio of this
source is
therefore real and not a spurious result of using non-simultaneous data of
a variable object.

It is apparent that \wga  looks like a High Energy Peaked source, with an
apparent
peak around 10$^{15.5}-10^{16.5} Hz$. The optical part of the SED at the
epoch of the SWIFT
observation looks parallel to that obtained from historical data (2MASS +
GSC2). We derived a
$B-R$ color index from the published
optical spectra in Caccianiga et al. (2000) and Perlman et al. (1998),
getting in both cases
$B-R\sim$0.47. This value
is in fair agreement with that derived extrapolating to the $R$ frequency
the UBV slope of the
SWIFT data ($B-R$=0.6).

\section{Discussion}

 At a redshift of 0.774 the
absolute $B$ magnitude  of \wga is -26.6 (H$_0$=70 q$_0$=0.5) and the radio
power at 1.4 GHz,
K-corrected assuming a slope for the radio flux density of -0.34, is 26.0
(Log W/Hz).
The source has therefore a radio power intermediate between those of FR I
and FR II
sources and is moderately radio-loud (F$_{radio}$/F$_{opt}$ $\sim$ 40). It
is reported as
unresolved with the VLA (Landt et al. 2006).

We compared the rms amplitude of the optical variability of \wga (0.27 mag)
with the corresponding value reported for the PG QSO sample (Giveon et al.
1999), derived from
light curves covering a time interval of 7 about 
years. No systematic difference between radio-loud and
radio-faint QSOs were found by Giveon et al. (1999) for this value, which
ranges from 0.05 to
0.34 mag:
the detected variability of \wga is therefore within the range of the PG
QSOs.

A much wider sample of optically selected QSO from the Sloan Digital Sky
Survey (SDSS), but
monitored only over a few year time base, by Vanden Berk et al. (2004) also
reports no definite
statistical difference between radio loud and radio quiet sources, while
the typical rms
variability is 0.13 ($b$) mag .

A comparison with longer light curves can be made using the Rosemary Hill
Observatory data
(Pica
et al. 1988) monitoring 144 AGNs (26 BL Lac, 18 radio quiet QSO, 85
radio-loud QSO and a few
miscellaneneus) over 19 years. 
Also in this case the rms variability of \wga is within the range of the
monitored sources and
also
the peak to peak variability amplitude (1.1 mag) is at an intermediate
level.

The X-ray brightness of \wga has therefore no apparent impact on its
optical variability.

In our opinion the historic light curve of \wga may be seen as
characterized by
a flat behaviour around B=16.0 with small(0.3 mag) amplitude oscillations
until JD
46500, when a fast jump down to 16.6 happened (see Fig.2). 
This vision is suggested by the fact that the POSS-I point at JD 35455 is
at the
average level of the light curve between JD 38000-46000, while the UVOT and
Ekar points are at
the level of the second part of the light curve (JD 46500-48000). However,
a simpler
interpretation
of a slow monotonic decreasing trend with
small amplitude oscillations cannot be ruled out.

The light curve shapes shown by Pica et al. (1988) are broadly classified
by them in three
classes: I) fast flickering with a nearly stable base level; II) small
flickering above a
long-term (possibly oscillating) trend; III)  significant flickering over
much slower long
term
changes of similar amplitude.
These classes are not sharply separated and they report some cases of
sources changing from
one
class to another. Two cases of "bistable" sources are also reported by
them, i.e. flickering
for some years above a given level and then, after a sharp transition,
above a different
level.
One is GC 0109+220 a well established BL Lac, the other is PKS 0723-008
which is presently
classified as a Narrow Line Radio Galaxy (Eracleous \& Halpern 2004) .
The recent behaviour  (1992-2002) of GC 0109+224 has been intensively
monitored by Ciprini et
al.
(2004): its behaviour may still be classified as a class I in the previous
scheme, but with a
higher base level than in the 1970's. From all the available data \wga
might belong to this
"bistable" class: clearly a longer monitoring would be necessary to check
this result.

\section{Conclusions}

Old photographic archives still contain a lot of unexplored data, which can
be very useful to
complement recent multi wavelength observations of variable sources (Nesci
et al. 2006). The optical light curve, spanning about
29 years, showed a moderate variability, typical for a QSO,
without strong flares (it is not an OVV object). The high X-ray/optical
flux ratio looks therefore real, and not 
a spurious result due to the use of non-simultaneous data.

A moderate level of variability is often found also in BL
Lacertae objects of the High Energy Peak type: \wga is not a BL Lac, given
its strong
Mg II 2900 \AA ~emission line with about 200 \AA ~of equivalent width, but
is a High Energy Peak object.

If a strong and fast variability is due to the fast cooling of relativistic
electrons emitting
by the synchrotron process, then it is expected to be observed mainly at
energies higher than the
peak of the SED (see  e.g. Perlman et al 2005): the mild behaviour of \wga
at optical frequencies
would therefore be quite within the current model expectations.

\begin{acknowledgments}
We thank the Directorate of the Asiago Observatory for hospitality during
the plates
scannings and telescope time allocation, and D. Trevese for helpful
suggestions. We thank also an anonimous referee for useful hints.
 
Part of this work was performed with the financial support of the 
Italian MIUR (Ministero dell' Istruzione, Universit\`a e Ricerca) 
under the grants Cofin 2002/024413 and 2003/027534.

The Guide Star Catalogue-II is a joint project of the Space Telescope
Science Institute and the Osservatorio Astronomico di Torino. Space
Telescope Science Institute is operated by the Association of
Universities for Research in Astronomy, for the National Aeronautics
and Space Administration under contract NAS5-26555. The participation
of the Osservatorio Astronomico di Torino is supported by the Italian
Council for Research in Astronomy. Additional support is provided by
European Southern Observatory, Space Telescope European Coordinating
Facility, the International GEMINI project and the European Space
Agency Astrophysics Division.

This research made use of the CDS (Strasbourg) and NED (NASA /IPAC
Extragalactic Database) databases.
\end{acknowledgments}


\clearpage

\begin{deluxetable}{lllcccc}
\tabletypesize{\footnotesize}
\tablecolumns{7}
\tablewidth{0pc}
\tablecaption{Reference stars in the field of \wga.
\label{std0447}}
\tablehead{
\colhead{RA(2000)}& \colhead{DEC(2000)}  & \colhead{GSC2} & \colhead{$R$
GSC2} &
\colhead{$B$ GSC2}& \colhead{$B$ Asiago} & \colhead{ident.}\\}
\startdata
\hline\\
04 47 29.672& -03 18 29.99& S020101354   &  14.65  &15.56 & 15.44 $\pm$
0.06& o \\
04 47 39.666& -03 18 27.48& S020101353   &  14.52  &15.56 & 15.48 $\pm$
0.05& n \\
04 47 42.546& -03 17 14.97& S020101350   &  14.87  &15.81 & 15.85 $\pm$
0.06& y \\
04 47 44.998& -03 21 21.27& S02010137258 &  15.67  &16.62 & 16.58 $\pm$
0.08& p \\
04 47 45.326& -03 22 41.42& S020101366   &  13.71  &14.76 & 14.86 $\pm$
0.06& k \\
04 47 46.044& -03 24 14.78& S02010137110 &  15.19  &16.09 & 16.00 $\pm$
0.07& r \\
04 47 47.440& -03 23 50.83& S020101372   &  14.36  &15.46 & 15.49 $\pm$
0.06& q \\
04 47 56.038& -03 26 15.88& S020101381   &  14.37  &15.32 & 15.28 $\pm$
0.06& j \\
04 47 59.969& -03 16 37.23& S02010119203 &  15.81  &16.74 & 16.68 $\pm$
0.10& b \\
04 48 03.515& -03 17 31.13& S02010119142 &  16.08  &16.78 & 16.69 $\pm$
0.10& z \\
04 48 04.352& -03 26 32.06& S020101385   &  13.27  &14.31 & 14.40 $\pm$
0.05& s \\
04 48 08.896& -03 27 36.46& S020101393   &  13.90  &15.03 & 15.03 $\pm$
0.06& v \\
04 48 08.973& -03 28 07.46& S020101394   &  13.86  &14.84 & 14.92 $\pm$
0.05& x \\
04 48 11.340& -03 22 11.41& S0201011270  &  13.69  &14.72 & 14.68 $\pm$
0.05& e \\
04 48 11.703& -03 27 17.69& S02010136903 &  15.45  &16.29 & 16.30 $\pm$
0.07& t \\
04 48 14.386& -03 21 59.36& S0201011266  &  14.07  &15.09 & 15.06 $\pm$
0.04& f \\
04 48 14.481& -03 27 39.03& S02010136876 &  14.93  &16.09 & 16.31 $\pm$
0.08& u \\
04 48 16.103& -03 21 11.37& S02010118906 &  14.90  &16.10 & 16.29 $\pm$
0.12& g \\

\enddata
\end{deluxetable}

\clearpage

\begin{deluxetable}{llll}
\tabletypesize{\footnotesize}
\tablecolumns{4}
\tablewidth{0pc}
\tablecaption{Photometry of \wga with UVOT.
\label{uvotmag}}
\tablehead{
\colhead{Filter}& \colhead{Mag}  & \colhead{error} & \colhead{$\nu
F_{\nu}$  } 
}
\startdata
\hline\\
V  & 16.53 & 0.04 & -11.310\\
B  & 16.91 & 0.03 & -11.259\\
U  & 15.97 & 0.02 & -11.223\\
W1 & 15.65 & 0.02 & -11.089\\
M2 & 15.63 & 0.01 & -10.962\\
W2 & 15.84 & 0.01 & -10.997\\
\enddata
\end{deluxetable}

\clearpage
\begin{figure}
\epsscale{1.0}
\includegraphics[scale=0.85]{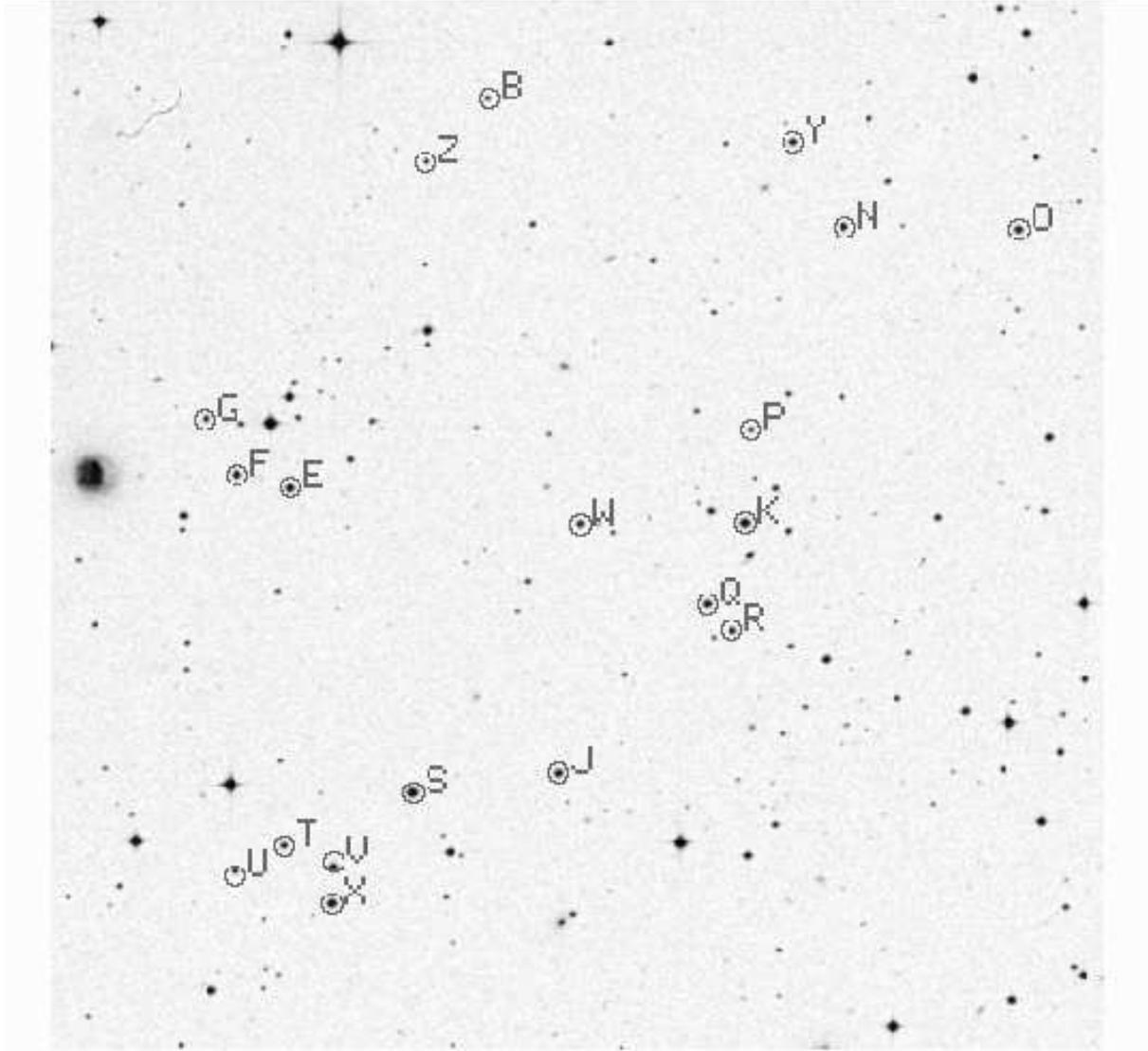}
\caption{Reference stars in the field of WGA 0447.9-0322. The source is
marked with 
W, North is up and East to the left, as usual. The image is from the
digitized plate  00031 of
the
S67 and the field of view is 15x15 arcmin.
\label{cartina}
}
\end{figure}

\clearpage

\begin{figure}
\epsscale{1.0} 
\includegraphics[scale=0.85]{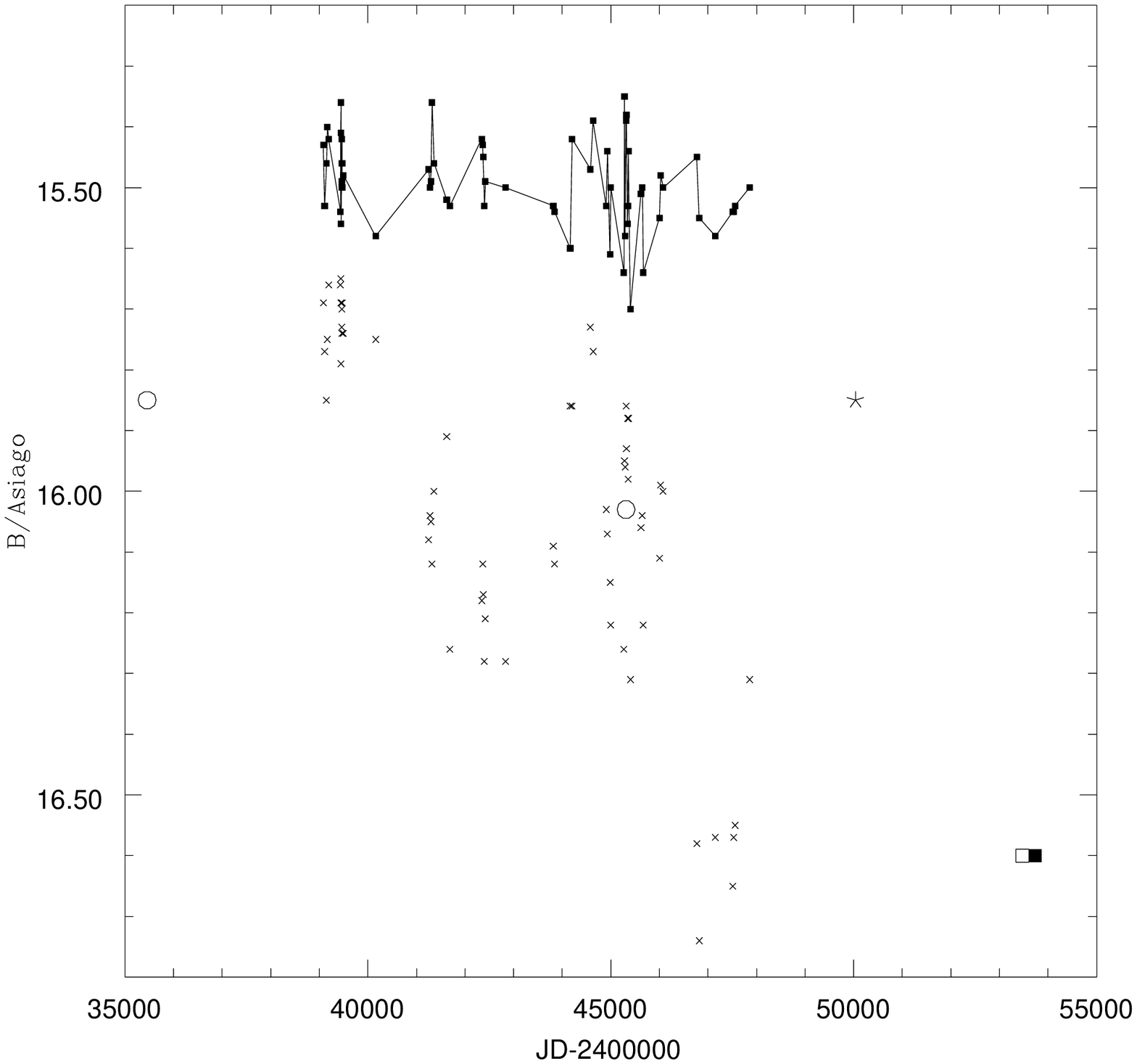}
\caption{Light curve of \wga from the 67cm Schmidt plates (crosses).
Abscissa is JD-2400000.
The light curve of the reference star r (filled squares, $B$=16.0) is
plotted for
comparison upshifted by 0.5 mag for clarity. Open circles are the POSS-I
(103aO) and the UKST
(IIIaJ) plate; star is
the UKST (IIIaF) plate converted to B/Asiago magnitude; open square is the
SWIFT/UVOT point
and filled square is the 
Ekar/AFOSC point  (see Sections 3 and 4 for details).
\label{luce67}
}
\end{figure}

\clearpage

\begin{figure}
\epsscale{1.0}
\includegraphics[scale=0.85]{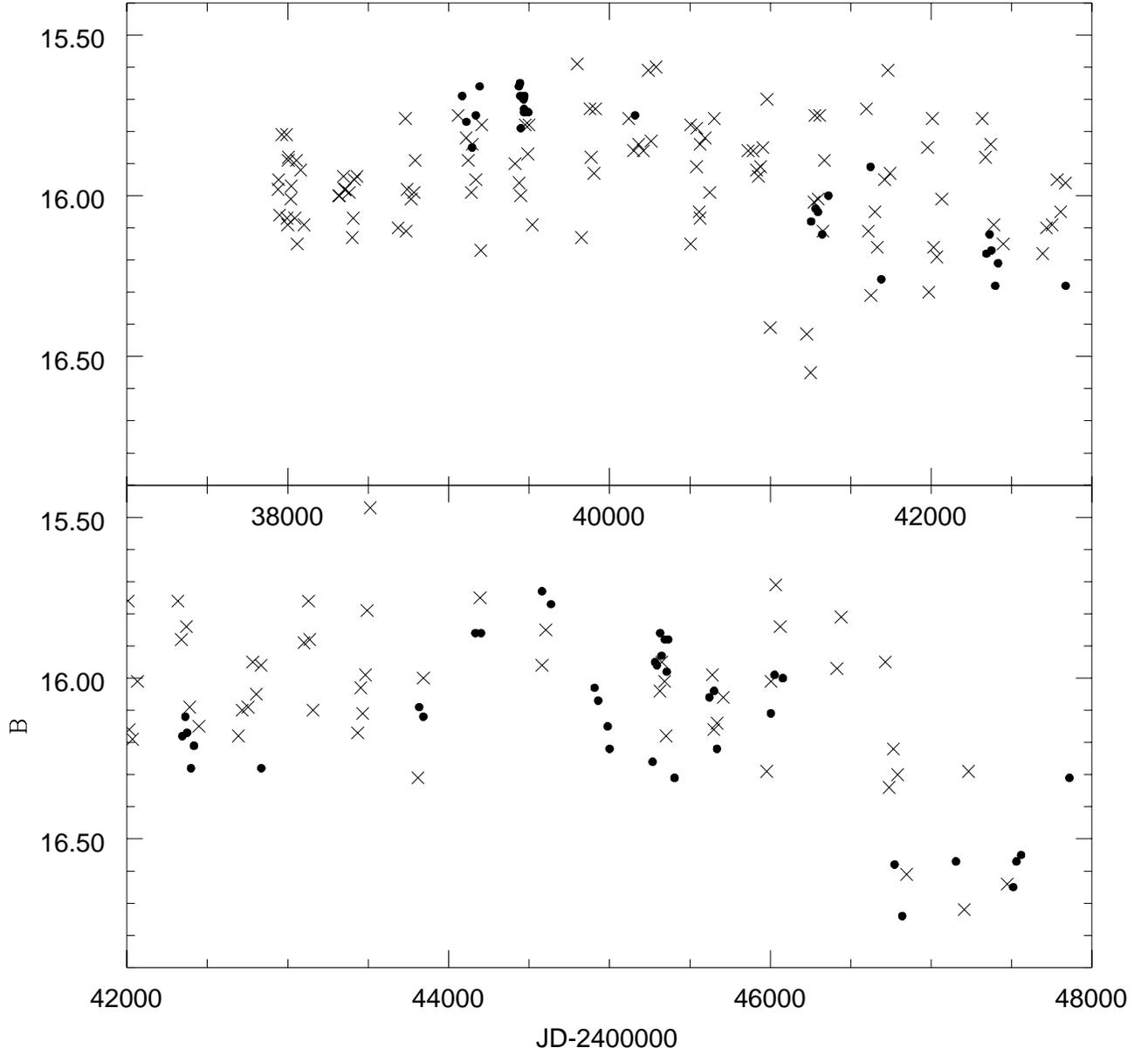}
\caption{The historic light curve in the $B$ band of \wga 
~from September 1962 to January 1991, using observations from both Asiago
Schmidt
telescopes.
Filled circles are S67 data and crosses the S40 ones.
\label{lc}
}
\end{figure}

\clearpage

\begin{figure}
\epsscale{1.0}
\includegraphics[scale=0.75]{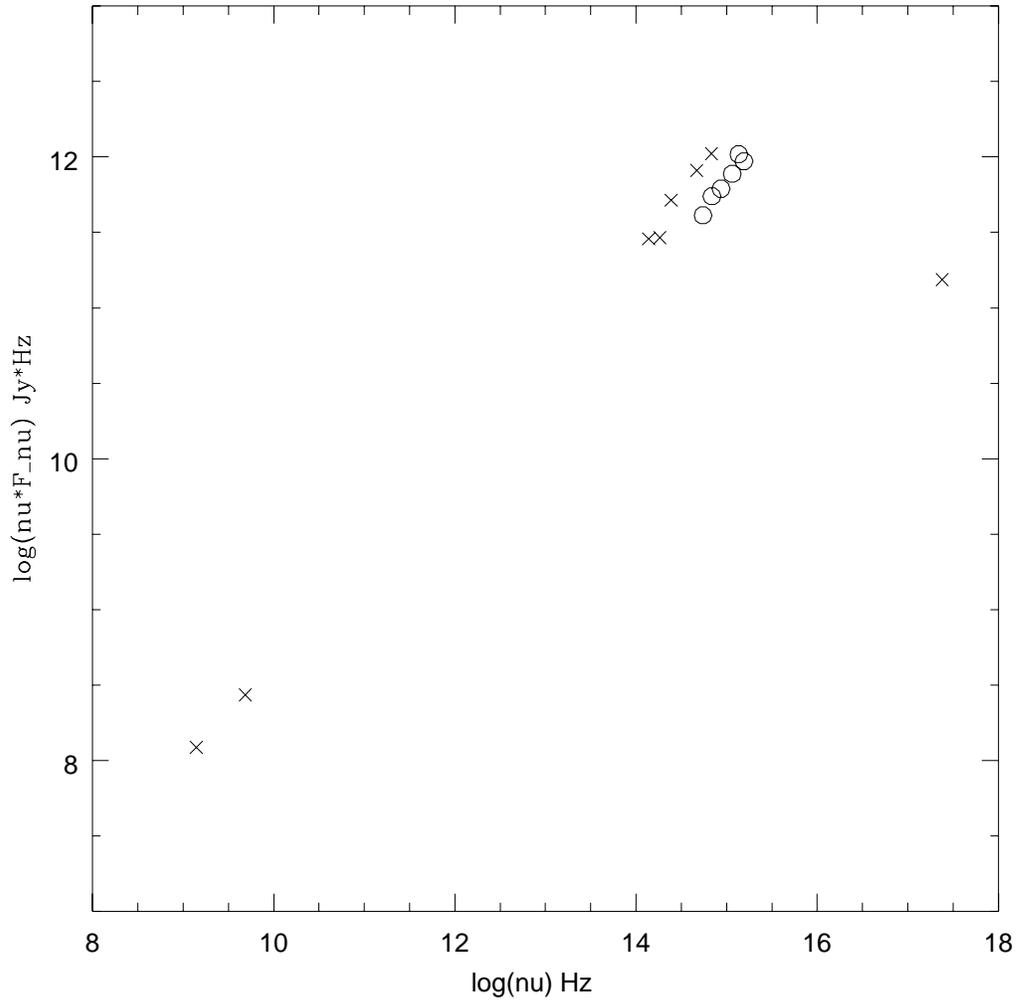}
\caption{The spectral energy distribution of \wga using literature data
(crosses)
and UVOT optical/UV data (circles). The X-ray point is from WGA (NED), JHK
from 2MASS, R,B from GSC2, radio data from NVSS and PMN. 
Note the fainter state of the source during the SWIFT pointing as compared
to the GSC2 (1982) epoch.
}
\end{figure}

\end{document}